\documentclass[%
 preprint,
 amsmath,amssymb,
 aps,
 pre,
showkeys
]{revtex4-2}

\usepackage{graphicx}
\usepackage{dcolumn}
\usepackage{bm}
\usepackage{hyperref}
\usepackage{xcolor}
\usepackage{ctable}

\begin{document}

\title{Transition temperature of a two-component Bose-Einstein condensates in improved Hartree-Fock approximation}

\author{Nguyen Van Thu$^*$}
\affiliation{Department of Physics, Hanoi Pedagogical University 2, Hanoi, Vietnam}
\email[]{nvthu@live.com}


\date{\today}

\begin{abstract}
In this study, we investigate the transition temperature of a two-component Bose-Einstein condensates by means of Cornwall–Jackiw–Tomboulis effective action formalism within the framework of the improved Hartree-Fock approximation. Influence of intra- and interspecies interactions as well as of the thermal fluctuations to the transition temperature are considered up to leading order of the gas parameters and scattering lengths.
\end{abstract}


\maketitle

\section{Introduction\label{sec1}}

One of the fundamental tasks in the study of Bose gas condensation is the determination of the transition temperature. The influence of interatomic interactions on this transition temperature has been a subject of extensive discussion and ongoing debate. In the case of a single weakly interacting Bose gas (BEC), while some studies have predicted a decrease in the transition temperature relative to that of an ideal Bose gas\cite{Toyoda1982,Wilkens2000}, numerous other works have reported an increase \cite{Grueter1997,Arnold2001a,Davis2003}. Both numerical simulations \cite{Kashurnikov2001,Arnold2001a} and recent theoretical analyses \cite{Thanh2024,VanThu2024} have elucidated that the repulsive interactions between bosonic atoms result in a positive relative shift in the transition temperature of a BEC.

For a mixture of Bose gases, in addition to intra-species interactions (i.e., interactions among atoms within the same species), inter-species interactions—those occurring between atoms of different species—also play a significant role \cite{Pitaevskij2010,Pethick2008}. These interactions lead to modifications of the transition temperature compared not only to that of an ideal Bose gas but also to that of a single-component BEC. To the best of our knowledge, the existing body of research on the transition temperature of two-component Bose–Einstein condensates (BECs) remains relatively limited. The first study addressing this issue was conducted by Shi {\it et.al.} \cite{Shi2000}, in which the transition temperatures of an immiscible two-component Bose–Einstein condensate (BEC) were analyzed across three distinct temperature regimes, each defined by the respective transition temperatures of the individual components. A subsequent investigation related to this topic is presented in Ref. \cite{Phat2009}, where the transition temperatures of the two components were determined by employing the Cornwall–Jackiw–Tomboulis (CJT) effective action formalism within the framework of the improved Hartree–Fock (IHF) approximation, through an analysis of the effective chemical potentials.

This paper is organised as follows. Section \ref{sec:2} focus on establishing the equations of motion of a dilute BECs. Section \ref{Ttrans} is to find the relative shift of transition temperature and thermodynamic quantities of a homogeneous dilute weakly interacting Bose gas by means of the CJT effective action approach at finite temperature. Finally, we present the conclusion and future outlook in Section \ref{conclusion}.

\section{Self-consistent equation of motion\label{sec:2}}

We begin by considering a BECs described by the Lagrangian density \cite{Pitaevskij2010,Pethick2008}
\begin{eqnarray}
{\cal L}=\sum_{j=1,2}\psi_j^*\left(-i\hbar\partial_t-\frac{\hbar^2}{2m_j}\nabla^2\right)\psi_j-V,\label{lagrangian}
\end{eqnarray}
where the Gross-Pitaevskii (GP) potential is given by
\begin{eqnarray}
V=\sum_{j=1,2}\left(-\mu_j|\psi_j|^2+\frac{G_{jj}}{2}|\psi_j|^4\right)+G_{12}|\psi_1|^2|\psi_2|^2,\label{potentialGP}
\end{eqnarray}  
Here, $\hbar$ denotes the reduced Planck constant. The mass and chemical potential of component $j$ are represented by $m_j$ and $\mu_j$, respectively. The field operator $\psi_j$, in general, is a function of both time $t$ and spatial coordinates $\vec{r}$. For a dilute BEC, the intra- and interspecies interactions are characterized by the coupling constants $G_{jj'}=2\pi\hbar^2a_{jj'}(1/m_j+1/m_{j'})$ with $a_{jj'}$ being the $s$-wave scattering length. In this study, we focus on the immiscible case, with repulsive interatomic interactions, i.e.,
\begin{eqnarray}
G_{11}G_{22}-G_{12}^2&>&0,\nonumber\\
a_{11},a_{22},a_{12}&>&0.\label{condition}
\end{eqnarray}

Let $\psi_{10}$ and $\psi_{20}$ denote the expectation values of the field operators. In the absence of external fields and particle flow, these quantities are real and serve as the order parameters, corresponding to the condensate densities $\rho_1=\psi_{10}^2,\rho_2=\psi_{20}^2$, respectively. Within the tree-level approximation, minimizing the GP potential (\ref{potentialGP})  with respect to the order parameters leads to the following gap equations
\begin{subequations}\label{gaptree}
  \begin{eqnarray}
  (-\mu_1+G_{11}\psi_{10}^2+G_{12}\psi_{20}^2)\psi_{10}&=&0,\label{gaptreea}\\
  (-\mu_2+G_{22}\psi_{20}^2+G_{12}\psi_{10}^2)\psi_{20}&=&0.\label{gaptreeb}
  \end{eqnarray}
\end{subequations} 
In the symmetry-broken phase for both components, the solutions to Eqs. (\ref{gaptree}) are
\begin{subequations}\label{psitree}
  \begin{eqnarray}
  \psi_{10}^2&=&\frac{\mu_1G_{22}-\mu_2G_{12}}{G_{11}G_{22}-G_{12}^2},\label{psitreea}\\
  \psi_{20}^2&=&\frac{\mu_2G_{11}-\mu_1G_{12}}{G_{11}G_{22}-G_{12}^2}.\label{psitreeb}
  \end{eqnarray}
\end{subequations} 

We now proceed to analyze the system within the Hartree-Fock approximation (HFA). To this end, the field operators are decomposed into fluctuations as follows   
\begin{eqnarray}
\psi_j\rightarrow \psi_{j0}+\frac{1}{\sqrt{2}}(\psi_{j1}+i\psi_{j2}),\label{shift}
\end{eqnarray}
where $\psi_{j1}$ and $\psi_{j2}$ represent the real and imaginary parts of the fluctuation fields, respectively. Inserting (\ref{shift}) into the Lagrangian (\ref{lagrangian}) one obtains the interacting Lagrangian density in two-loop approximation \cite{Phat2009,Thu2019,VanThu2022}
\begin{eqnarray}
{\cal L}_{\rm int}=&&\frac{1}{\sqrt{2}}\sum_{j=1,2}\left[G_{jj}\psi_{j0}\psi_{j1}+G_{12}\psi_{j'0}\psi_{j'1}\right](\psi_{j1}^2+\psi_{j2}^2)+\frac{1}{8}\sum_{j=1,2}G_{jj}(\psi_{j1}^2+\psi_{j2}^2)^2\nonumber\\
&&+\frac{G_{12}}{4}(\psi_{11}^2+\psi_{12}^2)(\psi_{21}^2+\psi_{22}^2).\label{Lint}
\end{eqnarray}
The propagator in tree-approximation can be written as
\begin{equation}
D_{j0}(k)=\frac{1}{\omega_n^2+E_{j0}^2(k)}\left(
              \begin{array}{cc}
               \varepsilon_{k,j} & \omega_n \\
                -\omega_n & \varepsilon_{k,j}+M_{j0}\\
              \end{array}
            \right),\label{eq:protree}
\end{equation}
in which $\varepsilon_{k,j}=\hbar^2k^2/2m_j$ is the kinetic energy of free particle. The Matsubara frequency $\omega_n$ is defined for boson as $\omega_n=2\pi n/\beta$ with $n\in{\mathbb{Z}}$. The effective mass is defined as 
\begin{eqnarray}
M_{j0}=G_{jj}\psi_{j0}^2.\label{masstree}
\end{eqnarray}
The energy spectrum can be found by examining pole of the propagator and in the tree-approximation (\ref{eq:protree}) gives
\begin{eqnarray}
E_{j0}(\vec{k})=\sqrt{\varepsilon_{k,j}\left(\varepsilon_{k,j}+M_{j0}\right)}.\label{disper0}
\end{eqnarray} 
In the long wavelength limit, Eq. (\ref{disper0}) shows a linear proportion in the wave vector
\begin{eqnarray}
E_{j0}\approx\sqrt{\frac{M_{j0}}{2m_j}}k,
\end{eqnarray}
which corresponds exactly to Goldstone bosons due to $U(1)\times U(1)$ breaking.

From the interaction Lagrangian (\ref{Lint}), the CJT effective potential $V_\beta^{\rm CJT}$ at finite temperature in the HFA can be read off \cite{Thu2019,VanThu2022},
\begin{eqnarray}
V_{\beta}^{\rm CJT}=&&\sum_{j=1,2}\left(-\mu_j|\psi_{j0}|^2+\frac{G_{jj}}{2}|\psi_{j0}|^4\right)+G_{12}|\psi_{10}|^2|\psi_{20}|^2\nonumber\\
&&+\frac{1}{2}\int_\beta\text{tr}\left\{\sum_{j=1,2}\left[\ln D_{j\rm(HFA)}^{-1}(k)+D_{j0}^{-1}(k)D_{j\rm(HFA)}(k)\right]-2.{1\!\!1}\right\}\nonumber\\
&&+\frac{3G_{11}}{8}(P_{11}^2+P_{22}^2)+\frac{G_{11}}{4}P_{11}P_{22}+\frac{3G_{22}}{8}(Q_{11}^2+Q_{22}^2)+\frac{G_{22}}{4}Q_{11}Q_{22}\nonumber\\
&&+\frac{G_{12}}{4}(P_{11}Q_{11}+P_{11}Q_{22}+P_{22}Q_{11}+P_{22}Q_{22}),\label{CJT1}
\end{eqnarray}
in which we abbreviate
\begin{eqnarray}
&&\int_\beta f(k)=\frac{1}{\beta}\sum_{n=-\infty}^{+\infty}\int \frac{d^3\vec{k}}{(2\pi)^3}f(\omega_n,\vec{k}),\nonumber\\
&&P_{aa}=\int_\beta D_{1\rm(HFA)aa},~Q_{aa}=\int_\beta D_{2\rm(HFA)aa},\label{momentum}
\end{eqnarray}
where $D_{j\rm(HFA)}(k)$ is the propagator in the HFA. Minimizing the CJT effective potential (\ref{CJT1}) with respect to the order parameters one arrives at the gap equations
\begin{eqnarray}
-\mu_1+G_{11}\psi_{10}^2+\frac{G_{12}}{2}\psi_{20}^2+\frac{G_{11}}{2}(3P_{11}+P_{22})+\frac{G_{12}}{4}(Q_{11}+Q_{22})&=&0,\nonumber\\
-\mu_2+G_{22}\psi_{20}^2+\frac{G_{12}}{2}\psi_{10}^2+\frac{G_{22}}{2}(3Q_{11}+Q_{22})+\frac{G_{12}}{4}(P_{11}+P_{22})&=&0.\label{gapHF}
\end{eqnarray}
Analogously, the propagators in the HFA are found by minimizing the CJT effective potential with respect to the elements of the propagators. The results are
\begin{eqnarray}
D_{1\rm(HFA)}(k)&=&\frac{1}{\omega_n^2+E_{1\rm(HFA)}^2(k)}\left(
              \begin{array}{cc}
                \varepsilon_{k,1}+\Pi_1^{(1)}& \omega_n \\
                -\omega_n & \varepsilon_{k,2}+\Pi_2^{(1)}\\
              \end{array}
            \right),\nonumber\\
D_{2\rm(HFA)}(k)&=&\frac{1}{\omega_n^2+E_{2\rm(HFA)}^2(k)}\left(
              \begin{array}{cc}
                \varepsilon_{k,2}+\Pi_1^{(2)}& \omega_n \\
                -\omega_n & \varepsilon_{k,2}+\Pi_2^{(2)}\\
              \end{array}
            \right),\nonumber\\
            \label{r12}
\end{eqnarray}
in which
\begin{eqnarray}
\Pi_1^{(1)}&=&-\mu_1+G_{11}\psi_{10}^2+\frac{G_{12}}{2}\psi_{20}^2+\frac{G_{11}}{2}(P_{11}+3P_{22})+\frac{G_{12}}{4}(Q_{11}+Q_{22}),\nonumber\\
\Pi_2^{(1)}&=&-\mu_1+3G_{11}\psi_{10}^2+\frac{G_{12}}{2}\psi_{20}^2+\frac{G_{11}}{2}(3P_{11}+P_{22})+\frac{G_{12}}{4}(Q_{11}+Q_{22}),\nonumber\\
\Pi_1^{(2)}&=&-\mu_2+G_{22}\psi_{20}^2+\frac{G_{12}}{2}\psi_{10}^2+\frac{G_{22}}{2}(Q_{11}+3Q_{22})+\frac{G_{12}}{4}(P_{11}+P_{22}),\nonumber\\
\Pi_2^{(2)}&=&-\mu_2+3G_{22}\psi_{20}^2+\frac{G_{12}}{2}\psi_{10}^2+\frac{G_{22}}{2}(3Q_{11}+Q_{22})+\frac{G_{12}}{4}(P_{11}+P_{22}).\label{SDHF}
\end{eqnarray}
Similarly to that in the tree-approximation, poles of the propagators (\ref{r12}) together with the gap equation (\ref{gapHF}) show a gap in energy spectrum of the excitations, which is equivalent to a violation of the Goldstone theorem. This situation also happens in Hartree-Fock-Bogoliubov theory, in which a gap in energy spectrum caused by the normal and anomalous densities \cite{Shi1998}.           However, it is well-known that the energy spectrum is gapless as proved for the BEC at zero temperature by Hugenholtz-Pines \cite{Hugenholtz1959} and later expanded for all values of temperature by Hohenberg  and Martin \cite{Hohenberg1965}.  In order to restore these phonons, the method developed in \cite{Ivanov2005} is employed. According to it, an extra term $\Delta V_\beta^{\rm CJT}$ needs to be added to the effective potential (\ref{CJT1}). As pointed out in Refs. \cite{Thu2019,VanThu2022},
\begin{eqnarray}
\Delta V_\beta^{\rm CJT}=&&-\frac{G_{11}}{4}(P_{11}^2+P_{22}^2)-\frac{G_{22}}{4}(Q_{11}^2+Q_{22}^2)\nonumber\\
&&+\frac{G_{11}}{2}P_{11}P_{22}+\frac{G_{22}}{2}Q_{11}Q_{22},
\end{eqnarray}
and thus one arrives at a new CJT effective potential 
\begin{eqnarray}
\widetilde{V}_\beta^{\rm CJT}=&&\sum_{j=1,2}\left(-\mu_j|\psi_{j0}|^2+\frac{G_{jj}}{2}|\psi_{j0}|^4\right)+G_{12}|\psi_{10}|^2|\psi_{20}|^2\nonumber\\
&&+\frac{1}{2}\int_\beta\mbox{tr}\left\{\sum_{j=1,2}\left[\ln D_j^{-1}(k)+D_{j0}^{-1}(k)D(k)\right]-2.{1\!\!1}\right\}+\frac{G_{11}}{8}(P_{11}^2+P_{22}^2)\nonumber\\
&&+\frac{3G_{11}}{4}P_{11}P_{22}+\frac{G_{22}}{8}(Q_{11}^2+Q_{22}^2)+\frac{3G_{22}}{4}Q_{11}Q_{22}\nonumber\\
&&+\frac{G_{12}}{4}(P_{11}Q_{11}+P_{11}Q_{22}+P_{22}Q_{11}+P_{22}Q_{22}).\label{CJT2}
\end{eqnarray}
This approximation is called the improved Hartree-Fock approximation (IHFA) and $D_j(k)$ is the propagator in this approximation. From this effective potential:

- Minimizing this effective potential with respect to the elements of the propagators one obtains the Schwinger-Dyson (SD) equations
\begin{eqnarray}
M_1&=&-\mu_1+3G_{11}\psi_{10}^2+G_{12}\psi_{20}^2+\Sigma_1^{(1)}\nonumber\\
M_2&=&-\mu_2+3G_{22}\psi_{20}^2+G_{12}\psi_{10}^2+\Sigma_1^{(2)}.\label{SD}
\end{eqnarray}
Here we use notations
\begin{eqnarray}
\Sigma_1^{(1)}&=&\frac{1}{2}(G_{11}P_{11}+3G_{11}P_{22}+G_{12}Q_{11}+G_{12}Q_{22}),\nonumber\\
\Sigma_1^{(2)}&=&\frac{1}{2}(G_{22}Q_{11}+3G_{22}Q_{22}+G_{12}P_{11}+G_{12}P_{22}).\label{mass}
\end{eqnarray}

- Minimizing this effective potential with respect to the order parameters leads to the gap equations
\begin{eqnarray}
-\mu_1+G_{11}\psi_{10}^2+G_{12}\psi_{20}^2+\Sigma_2^{(1)}&=&0,\nonumber\\
-\mu_2+G_{22}\psi_{20}^2+G_{12}\psi_{10}^2+\Sigma_2^{(2)}&=&0,\label{gap}
\end{eqnarray}
where
\begin{eqnarray}
\Sigma_2^{(1)}&=&\frac{1}{2}(3G_{11}P_{11}+g_{11}P_{22}+G_{12}Q_{11}+G_{12}Q_{22}),\nonumber\\
\Sigma_2^{(2)}&=&\frac{1}{2}(3G_{22}Q_{11}+g_{22}Q_{22}+G_{12}P_{11}+G_{12}P_{22}).\label{mass2}
\end{eqnarray}

- Combining the above, the inverse propagators have the form
\begin{eqnarray}
D_{j}(k)=\frac{1}{\omega_n^2+E_j^2(k)}\left(
              \begin{array}{lr}
                \varepsilon_{k,j} & \omega_n \\
                -\omega_n & \varepsilon_{k,j}+M_{j} \\
              \end{array}
            \right),\label{proIHF}
\end{eqnarray}
and the dispersion relation in the IHF approximation has the form
\begin{eqnarray}
E_j(k)=\sqrt{\varepsilon_{k,j}\left(\varepsilon_{k,j}+M_j\right)}.\label{dispIHF}
\end{eqnarray}
It is evident that the Goldstone bosons are correctly restored. The CJT effective potential, along with the SD equation (\ref{SD}) and the gap equations (\ref{gap}), satisfies the necessary conditions to ensure that the Goldstone theorem is upheld in the phase of spontaneously broken symmetry. Furthermore, these equations do not alter the HFA equations for the mean fields, as required in Ref. \cite{Ivanov2005}. This consistency is the basis for referring to the approach as the IHFA. The SD and gap equations are so-called the equation of motion.

At this stage, we consider the requirement of self-consistency in the equation of motion. To achieve this, Eqs. (\ref{SD}) and (\ref{gap}) are examined in conjunction with the effective potential presented in equation (\ref{CJT2}). Under these conditions, the CJT effective potential simplifies to
\begin{eqnarray}
\widetilde{V}_\beta^{\rm CJT}=&&\sum_{j=1,2}\left(-\mu_j|\psi_{j0}|^2+\frac{G_{jj}}{2}|\psi_{j0}|^4\right)+G_{12}|\psi_{10}|^2|\psi_{20}|^2\nonumber\\
&&+\frac{1}{2}\int_\beta\mbox{tr}\left[\sum_{j=1,2}\ln D_j^{-1}(k)\right]-\frac{G_{11}}{8}(P_{11}^2+P_{22}^2)-\frac{3G_{11}}{4}P_{11}P_{22}\nonumber\\
&&-\frac{G_{22}}{8}(Q_{11}^2+Q_{22}^2)-\frac{3G_{22}}{4}Q_{11}Q_{22}\nonumber\\
&&-\frac{G_{12}}{4}(P_{11}Q_{11}+P_{11}Q_{22}+P_{22}Q_{11}+P_{22}Q_{22}).\label{CJT3}
\end{eqnarray}
The pressure of the system is defined as minus of the CJT effective potential taking at the its minimum
\begin{eqnarray}
{\cal P}=-\widetilde V_\beta^{\rm CJT}\bigg|_{\rm minimum}.\label{pressure}
\end{eqnarray}
The condition of the self-consistence requires the chemical potential equals the first derivative of the pressure with respect to the particle density
\begin{eqnarray}
\mu_j=\frac{\partial \cal P}{\partial \rho_j}.\label{chemical}
\end{eqnarray}
From Eqs. (\ref{CJT3})-(\ref{chemical}) one finds the chemical potentials in the IHFA
\begin{eqnarray}
\mu_1&=&\mu_{10}+G_{11}P_{11},\nonumber\\
\mu_2&=&\mu_{20}+G_{22}Q_{11},\label{muIHFA}
\end{eqnarray}
where $\mu_{j0}$ is the chemical potential at zero temperature of the component $j$. Plugging (\ref{muIHFA}) into (\ref{SD}) and (\ref{gap}) one arrives at

- The SD equations
\begin{eqnarray}
M_1&=&-\mu_{10}+3G_{11}\psi_{10}^2+G_{12}\psi_{20}^2+\frac{1}{2}(-G_{11}P_{11}+3G_{11}P_{22}+G_{12}Q_{11}+G_{12}Q_{22}),\nonumber\\
M_2&=&-\mu_{20}+3G_{22}\psi_{20}^2+G_{12}\psi_{10}^2+\frac{1}{2}(-G_{22}Q_{11}+3G_{22}Q_{22}+G_{12}P_{11}+G_{12}P_{22}).\label{SDself}
\end{eqnarray}

- The gap equations
\begin{eqnarray}
-\mu_{10}+G_{11}\psi_{10}^2+G_{12}\psi_{20}^2+\frac{1}{2}(G_{11}P_{11}+G_{11}P_{22}+G_{12}Q_{11}+G_{12}Q_{22})&=&0,\nonumber\\
-\mu_{20}+G_{22}\psi_{20}^2+G_{12}\psi_{10}^2+\frac{1}{2}(G_{22}Q_{11}+G_{22}Q_{22}+G_{12}P_{11}+G_{12}P_{22})&=&0.\label{gapself}
\end{eqnarray}
A notable property, as observed from Eqs. (\ref{masstree}) and (\ref{SDself}), is that within the IHFA, the effective masses depend on both intra- and interspecies interactions, whereas in the tree-level approximation, they depend solely on the intraspecies interactions. This ambiguity is removable if we rewrite  Eqs. (\ref{SDself}) and (\ref{gapself}) in form
\begin{eqnarray}
M_1&=&2G_{11}\psi_{10}^2-G_{11}(P_{11}-P_{22}),\nonumber\\
M_2&=&2G_{22}\psi_{20}^2-G_{22}(Q_{11}-Q_{22}).\label{Mself}
\end{eqnarray}
Equations (\ref{gapself}) and (\ref{Mself}) are referred to as the self-consistent equations of motion, which govern the evolution of the system.

\section{Transition temperatures\label{Ttrans}} 

In a BEC mixture, phase segregation is strongly depends by intrinsic atomic parameters and temperature. Let $T_{C1}$ and $T_{C2}$ denote the critical transition temperatures of components 1 and 2, respectively, with the assumption that $T_{C1}< T_{C2}$. Depending on the value of the system temperature $T$ relative to these transition temperatures, the system may exhibit the following three distinct regimes:

- Low-temperature regime ($T<T_{C1}$): Both components exist in the condensed phase.

- Intermediate-temperature regime ($T_{C1}<T<T_{C2}$): Component 1 undergoes a transition to the normal phase, while component 2 remains in the condensed phase.

- High-temperature regime ($T>T_{C2}$): Both components are found in the normal phase.

In order to investigate these transition temperatures we now consider these regimes.

\subsection{Low-temperature regime ($T<T_{C1}$)}

When the temperature is below $T_{C1}$, both components are considered to be in the condensed phase; more precisely, a number of atoms from each component exist in the condensed phase.
To facilitate the investigation of transition temperatures, it is essential to first evaluate the momentum integrals $P_{11},P_{22},Q_{11}$ and $Q_{22}$, which arise in the SD and gap equations. These integrals are defined in Eq. (\ref{momentum}) within the framework of the HFA. In the context of the IHFA, the definitions remain applicable, with the sole modification being the substitution of the HFA propagator with that of the IHFA, as specified in Eq.  (\ref{proIHF}).

It is not difficult to elucidate that in high temperature limit these momentum integrals can be written as \cite{VanThu2024}
\begin{eqnarray}
P_{11}&=&\frac{(2m_1)^{3/2}M_1^{3/2}}{6\pi^2\hbar^3}+\frac{\zeta(3/2)}{\lambda_1^3}-\frac{m_1\zeta(1/2)M_1}{4\pi\hbar^2\lambda_1},\nonumber\\
P_{22}&=&-\frac{(2m_1)^{3/2}M_1^{3/2}}{12\pi^2\hbar^3}+\frac{\zeta(3/2)}{\lambda_1^3}+\frac{3m_1\zeta(1/2)M_1}{4\pi\hbar^2\lambda_1}\nonumber\\
Q_{11}&=&\frac{(2m_2)^{3/2}M_2^{3/2}}{6\pi^2\hbar^3}+\frac{\zeta(3/2)}{\lambda_2^3}-\frac{m_2\zeta(1/2)M_2}{4\pi\hbar^2\lambda_2},\nonumber\\
P_{22}&=&-\frac{(2m_2)^{3/2}M_2^{3/2}}{12\pi^2\hbar^3}+\frac{\zeta(3/2)}{\lambda_2^3}+\frac{3m_2\zeta(1/2)M_2}{4\pi\hbar^2\lambda_2},
\label{chottichphan}
\end{eqnarray}
in which the de Broglie wave length is defined as $\lambda_j=\sqrt{2\pi\hbar^2/m_jk_BT}$ and zeta function has the form $\zeta(x)=\sum_{n=1}^\infty 1/n^x$.

To proceed with the analysis, we introduce the following dimensionless quantities: the effective mass ${\cal M}_j=M_j/G_{jj}\rho_j$, the gas parameter $\alpha_j=\rho_ja_{jj}^3$, and the reduced order parameter 
$\phi_{j0}=\psi_{j0}/\sqrt{\rho_j}$. The system under consideration consists of dilute BECs, which implies that the scattering lengths are not only much smaller than the average distance between bosonic atoms but also significantly smaller than the corresponding de Broglie wavelength, i.e., $\alpha_j\ll1$ and $t_j\equiv a_{jj}/\lambda_j\ll1$ \cite{Andersen2004,Shi1998}. This reflects the dilute nature of the condensates, as well as the weak interatomic interactions within the system. In this regard, substituting (\ref{chottichphan}) into (\ref{Mself}) yields
\begin{eqnarray}
{\cal M}_1&=&2f_1-\frac{4\sqrt{2}\alpha_1^{1/2}}{\pi^{1/2}}{\cal M}_1^{3/2}+\frac{4\zeta(1/2)a_{11}}{\lambda_1}{\cal M}_1,\nonumber\\
{\cal M}_2&=&2f_2-\frac{4\sqrt{2}\alpha_1^{1/2}}{\pi^{1/2}}{\cal M}_2^{3/2}+\frac{4\zeta(1/2)a_{22}}{\lambda_1}{\cal M}_2,\label{Mvsf}
\end{eqnarray}
in which we have defined the condensed fraction of component $j$ as $f_j=\rho_{j0}/\rho_j$ with $\rho_j$ being the total particle number of $j$-component. 
To leading order in the small parameters, Eqs. (\ref{Mvsf}) allow us to express the dimensionless effective masses as functions of the condensed fractions
\begin{eqnarray}
{\cal M}_1&=&2f_1+\frac{8\zeta(1/2)a_{11}}{\lambda_1}f_1-\frac{16\alpha_1^{1/2}}{\pi^{1/2}}f_1^{3/2},\nonumber\\
{\cal M}_2&=&2f_2+\frac{8\zeta(1/2)a_{22}}{\lambda_2}f_2-\frac{16\alpha_2^{1/2}}{\pi^{1/2}}f_2^{3/2}.\label{Mvsf1}
\end{eqnarray}
To process further, we introduce two dimensionless control parameters
\begin{eqnarray}
K_1=\frac{(m_1+m_2)n_1a_{12}}{2m_1n_2a_{22}},~K_2=\frac{(m_1+m_2)n_2a_{12}}{2m_2n_1a_{11}}.\label{2K}
\end{eqnarray}
The condition of immiscibility necessitates that $K_1,K_2>1$. The relationship between $K_1$ and $K_2$ is established by imposing the constraint that the bulk pressures of the two bulk phases are equal
\begin{eqnarray}
\frac{\mu_1^2}{2G_{11}}=\frac{\mu_2^2}{2G_{22}}.\label{equal}
\end{eqnarray}
In terms of (\ref{Mvsf1}) and (\ref{2K}), the gap equations (\ref{gapself}) are approximated as
\begin{eqnarray}
&-&\tilde\mu_{10}+f_1+2t_1\zeta(1/2)f_1+\frac{8\alpha_1^{1/2}f_1^{3/2}}{3\sqrt{\pi}}+\frac{\zeta(3/2)}{\rho_1\lambda_1^3}\nonumber\\
&+&K_2f_2+2K_2t_2\zeta(1/2)f_2+\frac{8K_2\alpha_2^{1/2}f_2^{3/2}}{3\sqrt{\pi}}
+\frac{K_2\zeta(3/2)}{\rho_2\lambda_2^3}=0,\label{gap1}
\end{eqnarray}
and
\begin{eqnarray}
&-&\tilde\mu_{20}+f_2+2t_2\zeta(1/2)f_2+\frac{8\alpha_2^{1/2}f_2^{3/2}}{3\sqrt{\pi}}+\frac{\zeta(3/2)}{\rho_2\lambda_2^3}\nonumber\\
&+&K_1f_1+2K_1t_1\zeta(1/2)f_1+\frac{8K_1\alpha_2^{1/2}f_1^{3/2}}{3\sqrt{\pi}}
+\frac{K_1\zeta(3/2)}{\rho_1\lambda_1^3}=0,\label{gap2}
\end{eqnarray}
where the dimensionless chemical potential at zero temperature is defined as $\tilde\mu_{j0}=\mu_{j0}/G_{jj}\rho_j$ and $t_j=a_{jj}/\lambda_j$.

The first case to be considered is when quantum fluctuations are neglected. The solutions to the gap equations (\ref{gap1}) and (\ref{gap2}) are as follows
\begin{eqnarray}
f_1^{\rm(0)}&=&\frac{1}{1+2\zeta(1/2)t_1}\left[\frac{K_2\tilde\mu_{20}-\tilde\mu_{10}}{K_1K_2-1}-\frac{\zeta(3/2)}{\rho_1\lambda_1^3}\right],\nonumber\\
f_2^{\rm(0)}&=&\frac{1}{1+2\zeta(1/2)t_2}\left[\frac{K_1\tilde\mu_{10}-\tilde\mu_{20}}{K_1K_2-1}-\frac{\zeta(3/2)}{\rho_2\lambda_2^3}\right].\label{flow}
\end{eqnarray}
At the transition temperature the condensed fraction vanishes, Eq. (\ref{flow}) leads
\begin{eqnarray}
T_{C1}&=&\frac{2\pi\hbar^2}{m_1k_B}\left[\frac{\rho_1}{\zeta(3/2)}\right]^{2/3}\left(\frac{K_2\tilde\mu_{20}-\tilde\mu_{10}}{K_1K_2-1}\right)^{2/3},\nonumber\\
T_{C2}&=&\frac{2\pi\hbar^2}{m_2k_B}\left[\frac{\rho_2}{\zeta(3/2)}\right]^{2/3}\left(\frac{K_1\tilde\mu_{10}-\tilde\mu_{20}}{K_1K_2-1}\right)^{2/3}.\label{Tc1}
\end{eqnarray}
Neglecting the quantum fluctuations is equivalent to the vanish of the last terms in right-hand side of Eq. (\ref{muIHFA}). In this case one gas
\begin{eqnarray}
\mu_{10}=G_{11}\rho_1+G_{12}\rho_2,\nonumber\\
\mu_{20}=G_{22}\rho_2+G_{12}\rho_1.\label{mu0}
\end{eqnarray}
Combining (\ref{mu0}) with (\ref{2K}) and (\ref{equal}), it is easy to see that $K_1=K_2=K$, therefore  the dimensionless chemical potentials can be expressed as
\begin{eqnarray}
\tilde\mu_{10}=\tilde\mu_{20}=1+K.\label{mu01}
\end{eqnarray}
 \cite{Phat2009,Son2002}. The result (\ref{Tc1}) reduces to
\begin{eqnarray}
T_{C1}^{\rm(0)}&=&\frac{2\pi\hbar^2}{m_1k_B}\left[\frac{\rho_1}{\zeta(3/2)}\right]^{2/3},\nonumber\\
T_{C2}^{\rm(0)}&=&\frac{2\pi\hbar^2}{m_2k_B}\left[\frac{\rho_2}{\zeta(3/2)}\right]^{2/3}.\label{Tc3}
\end{eqnarray}
These results indicate that the transition temperatures of dilute BECs are identical to those of ideal BECs \cite{Pitaevskij2010,Pethick2008}; that is, interatomic interactions do not affect the transition temperature. 

We shall now undertake a detailed examination of the influence exerted by interatomic interactions on the transition temperature, achieved through the incorporation of quantum fluctuations into the gap equations (\ref{gap1}) and (\ref{gap2}). It is well recognized that, in the case of dilute BECs, the contribution arising from quantum fluctuations is sufficiently negligible to warrant treatment as a perturbative correction. In this regard, the corresponding solutions to gap equations (\ref{gap1}) and (\ref{gap2}) are presented as follows
\begin{eqnarray}
f_1&=&f_1^{\rm(0)}-\frac{8\alpha_1^{1/2}}{3\sqrt{\pi}}\left[1-2\zeta(1/2)t_1\right]\left[\frac{K_2\tilde\mu_{20}-\tilde\mu_{10}}{K_1K_2-1}-\frac{\zeta(3/2)}{\rho_1\lambda_1^3}\right]^{3/2},\nonumber\\
f_2&=&f_2^{\rm(0)}-\frac{8\alpha_2^{1/2}}{3\sqrt{\pi}}\left[1-2\zeta(1/2)t_2\right]\left[\frac{K_1\tilde\mu_{10}-\tilde\mu_{20}}{K_1K_2-1}-\frac{\zeta(3/2)}{\rho_2\lambda_2^3}\right]^{3/2}.\label{flow1}
\end{eqnarray}
It is easy to see that at zero temperature and in strong segregation limit, i.e., $K_1,K_2\rightarrow\infty$, Eq. (\ref{flow1}) gives
\begin{eqnarray}
f_1&=&1-\frac{8\sqrt{\alpha_1}}{3\sqrt{\pi}},\nonumber\\
f_2&=&1-\frac{8\sqrt{\alpha_2}}{3\sqrt{\pi}}.\label{limit1}
\end{eqnarray}
Obviously that in the strong segregation limit, two components are independent. This well-known result was first theoretically discovered by Bogoliubov \cite{Bogoliubov1947} and subsequently reaffirmed by numerous authors in the context of a single BEC \cite{Wu1959,Carlen2021a,VanThu2022a}. More recently, it has also been confirmed through experimental observations \cite{Lopes2017}.

We now find the transition temperatures. Assume that the relative shifts of these transition temperature in compared with those in case ignoring the quantum fluctuations, the transition temperatures can be written in form
\begin{eqnarray}
T_{Cj}=T_{Cj}^{\rm(0)}(1+t).\label{approx}
\end{eqnarray}
Plugging (\ref{approx}) into (\ref{flow}) and requiring the vanishing of the condensed fractions we have
\begin{subequations}\label{Tcful}
\begin{eqnarray}
T_{C1}&=&\frac{2\pi\hbar^2}{m_1k_B}\left[\frac{\rho_1}{\zeta(3/2)}\right]^{2/3}\left[1-\frac{4\zeta(1/2)}{3\zeta(3/2)^{1/3}}\left(1-\frac{K_2\tilde\mu_{20}-\tilde\mu_{10}}{K_1K_2-1}\right)\rho_1^{1/3}a_{11}\right],\label{Tcfula}\\
T_{C2}&=&\frac{2\pi\hbar^2}{m_2k_B}\left[\frac{\rho_2}{\zeta(3/2)}\right]^{2/3}\left[1-\frac{4\zeta(1/2)}{3\zeta(3/2)^{1/3}}\left(1-\frac{K_1\tilde\mu_{10}-\tilde\mu_{20}}{K_1K_2-1}\right)\rho_2^{1/3}a_{22}\right].\label{Tcfulb}
\end{eqnarray}
\end{subequations}
In the strong segregated limit, Eq. (\ref{Tcful}) gives
\begin{subequations}\label{Tcstrong}
\begin{eqnarray}
T_{C1}&=&\frac{2\pi\hbar^2}{m_1k_B}\left[\frac{\rho_1}{\zeta(3/2)}\right]^{2/3}\left[1-\frac{4\zeta(1/2)}{3\zeta(3/2)^{1/3}}\rho_1^{1/3}a_{11}\right],\label{Tcstronga}\\
T_{C2}&=&\frac{2\pi\hbar^2}{m_2k_B}\left[\frac{\rho_2}{\zeta(3/2)}\right]^{2/3}\left[1-\frac{4\zeta(1/2)}{3\zeta(3/2)^{1/3}}\rho_2^{1/3}a_{22}\right].\label{Tcstrongb}
\end{eqnarray}
\end{subequations}
It is readily observed that the transition temperature in the strong segregation limit coincides with that of a single BEC, as established in Ref. \cite{VanThu2024}. Equation (\ref{Tcstrong}) further demonstrates that, within this limit, interspecies interactions exert no influence on the critical temperature.

\subsection{Intermediate-temperature regime}

In this regime, all atoms of component 1 reside in the normal phase, whereas a portion of atoms from component 2 remains in the condensed phase. This indicates that the condensed fraction $f_1$ of component 1 vanishes, while that of component 2 remains nonzero. Setting $f_1=0$ in the gap and SD equations (\ref{SDself}) and (\ref{gapself})
\begin{eqnarray}
K_1\tilde\mu_{10}-\tilde\mu_{20}-\frac{1}{2}(K_1K_2-1){\cal M}_2+(K_1K_2-1){\cal M}_2t_2\zeta(1/2)-(K_1K_2-1)\frac{\zeta(3/2)}{\rho_2\lambda_2^3}=0,\label{SD2t}
\end{eqnarray}
and
\begin{eqnarray}
f_2=\frac{1}{2}\left({\cal M}_2+\frac{Q_{11}}{\rho_2}-\frac{Q_{22}}{\rho_2}\right).\label{gap2t}
\end{eqnarray}
Solving Eqs. (\ref{SD2t}) and (\ref{gap2t}) one finds approximate solution for the condensed fraction of component 2
\begin{eqnarray}
f_2=\frac{K_1\tilde\mu_{10}-\tilde\mu_{20}}{K_1K_2-1}-\frac{\zeta(3/2)}{\rho_2\lambda_2^3}-\frac{2\zeta(1/2)a_{22}}{\lambda_2}\left[\frac{K_1\tilde\mu_{10}-\tilde\mu_{20}}{K_1K_2-1}-\frac{\zeta(3/2)}{\rho_2\lambda_2^3}\right].\label{f22}
\end{eqnarray}
At $T=T_{C2}$, all atoms of component 2 are in the normal phase, i.e., $f_2=0$. Eq. (\ref{f22}) gives
\begin{eqnarray}
T_{C2}=T_{C2}^{\rm(0)}\left[1-\frac{4(K_2m_2^{3/2}m_1^{-1/2}\rho_1-K_1m_1m_2\rho_2)^2\zeta(1/2)}{3(m_2^{3/2}m_1^{-1/2}\rho_1-K_1m_1m_2\rho_2)^2\zeta(3/2)^{1/3}}\rho_2^{1/3}a_{22}\right].\label{Tc22}
\end{eqnarray}
In the strong segregation, Eq. (\ref{Tc22}) reduces to (\ref{Tcstrongb}). 

To conclude this section, it is important to note that, at zero temperature and under the assumption that quantum fluctuations are negligible, all atoms of the components exist solely in the condensed phase. In order to determine the transition temperatures, we have examined the behavior of the condensed fractions across two distinct temperature regimes relative to the transition temperatures. In the case where $T_{C1}>T_{C2}$, the results remain consistent, with the indices 1 and 2 being interchangeable.

\section{Conclusion and outlook\label{conclusion}}

In the preceding sections, the transition temperatures of a two-component BECs have been investigated within the IHF approximation, employing the CJT effective action formalism. The principal findings of our study are as follows:

- The self-consistent equations of motion for a dilute two-component BEC have been derived within the CJT effective action framework, ensuring the preservation of the Goldstone modes arising from the spontaneous breaking of the $U(1)\times U(1)$ symmetry.

- Both intra-species and inter-species interactions among the bosonic atoms exert a significant influence on the transition temperatures.

- In the strong segregation limit, each component behaves effectively as an independent single-component BEC.

\begin{acknowledgements}
This research is funded by Vietnam National Foundation for Science and Technology Development (NAFOSTED) under grant number 103.01-2023.12.
\end{acknowledgements}

\section*{Conflict of interest}
All of the authors declare that we have no conflict of interest.

\bibliography{2becs.bib}

\end{document}